\documentclass[a4paper,twoside]{article}

\usepackage{epsfig}
\usepackage{subcaption}
\usepackage{calc}
\usepackage{amssymb}
\usepackage{amstext}
\usepackage{amsmath}
\usepackage{amsthm}
\usepackage{multicol}
\usepackage{pslatex}
\usepackage{url}
\usepackage{apalike}
\usepackage{algorithm2e}
\usepackage[bottom]{footmisc}
\usepackage{SCITEPRESS}

\begin{document}

\title{Block-Based Pathfinding: A Minecraft System for Visualizing Graph Algorithms}

\author{\authorname{Luca-Ștefan Pîrvu\sup{1}\orcidAuthor{0009-0005-5484-3796}, Bogdan-Alexandru Măciucă\sup{1}\orcidAuthor{0009-0009-3479-966X}, Andrei-Ciprian Râbu\sup{1}\orcidAuthor{0009-0005-2215-5438} and Adrian-Marius Dumitran\sup{1}\orcidAuthor{0009-0005--3547-5772}}
\affiliation{\sup{1}University of Bucharest, Faculty of Mathematics and Informatics, Academiei 14, 010014, Bucharest, Romania}
\email{luca.pirvu@gmail.com, bogdanmaciuca0621@gmail.com, andrei.ciprian.r@gmail.com, marius.dumitran@unibuc.ro}
}

\keywords{Game-Based Learning, Computer Science Education, Graph Theory, Dijkstra, A*, Virtual Environments.}

\abstract{
  Graph theory is a cornerstone of Computer Science education, yet entry-level students often struggle to map abstract node-edge relationships to practical applications. This paper presents the design and architecture of a Minecraft-based educational tool specifically built to visualize graph traversal and shortest-path algorithms. We propose a three-layer system: (1) a Grid Traversal module where terrain types (e.g., soul sand, ice) represent edge weights, allowing for the gamified study of shortest path algorithms; (2) a "Sky Graph" module for interactive 3D manipulation of both directed and undirected graphs; and (3) lessons and quizzes available through books. The system grounds its design in Constructionist learning theory, transitioning students from passive observers to active protagonists who physically manipulate algorithmic behavior. We additionally present a planned empirical evaluation using NASA-TLX and in-game telemetry to validate the system's pedagogical efficacy.
}

\onecolumn \maketitle \normalsize 
\setcounter{footnote}{0} \vfill

This is the author's version of the work. The definitive version was published in the Proceedings of the 18th International Conference on Computer Science and Education (CSEDU 2026)

\section{\uppercase{Introduction}}
\label{sec:introduction}

\noindent Graph theory concepts such as traversal, weighted paths, and acyclicity are standard topics in high school CS electives and freshman university courses. However, the transition from whiteboard diagrams to code implementation (or an adequate mental model) poses a significant cognitive load. Furthermore, it has been shown that the vast majority of students view learning activities inside a typical organized environment, such as a school or college classroom, as restrictive and boring.

As one of the most popular games, \textit{Minecraft} has nearly 32 million daily active players, 43\% of which are between the ages of 15 and 21 years old \cite{DemandSage}. Just by its core gameplay mechanics, players can learn a variety of skills, such as managing time and resources, calculating risk and reward ratios, circuit building (via the \textit{Redstone} mechanics) or spatial reasoning. The game is based on a virtual world which offers a native grid structure that extends a simple 2D array and thus is easy to conceptualize. While existing educational games effectively teach basic programming or foundational mathematics, there is a notable lack of interactive tools designed specifically for discrete mathematics and graph theory.

This paper proposes a system that leverages the \textit{Minecraft} environment to transform abstract graph algorithms and concepts into tangible, interactive challenges and lessons, opening the door to more freedom in experimentation and faster iteration. The sandbox nature of the game provides a high level of immersion, minimizing distractions and fostering focused problem-solving.

The main contributions of this paper are:
\begin{itemize}
\item The design and implementation of a custom \textit{Minecraft} modification (commonly referred to as a "mod") tailored for graph theory instruction.
\item A framework for translating abstract graph traversal algorithms into spatial, in-game challenges.
\end{itemize}

The remainder of this paper is organized as follows. Section 2 reviews related work in game-based learning and computer science education. Section 3 details the system design and the methodology for balancing educational content with gameplay. Section 4 provides an overview of the technical implementation and software architecture. Section 5 presents the pedagogical justification, Section 6 describes the planned empirical evaluation, Section 7 discusses future work, and Section 8 concludes the paper.

The supplementary materials for this work, including the underlying dataset and comparative analysis files, have been made available for peer review at the following anonymized URL.\footnote{\url{https://anonymous.4open.science/r/MinecraftMcMapsFabric}}

\section{\uppercase{Related Work}}
\label{sec:relatedwork}

\subsection{Algorithm Visualization and Game-Based Learning}
\noindent The persistent challenge in computer science education is maintaining student engagement, especially when transitioning from traditional classroom environments to digital platforms. Students demonstrate significant reluctance to utilize passive, non-interactive virtual learning environments, which often suffer from poor design and fail to foster active participation \cite{Arouri2024}. Abstract theoretical topics, such as automata and graph theory, present a high cognitive load, and students require interactive, puzzle-solving approaches to successfully bridge the gap between theoretical models and practical implementation \cite{Jordaan2025}.

Algorithm visualization has proven to be a highly effective pedagogical strategy for these abstract topics, with students explicitly requesting visual, step-by-step tools for complex data structures like graphs and trees \cite{Simonak2014}. Combining these visualizations with Game-Based Learning (GBL) yields statistically higher motivation levels, yet many custom-built 2D educational games are limited by fixed perspectives and restricted interactivity, which can hinder the exploration of multi-dimensional data structures \cite{Meissner2026}. To overcome these constraints, researchers have emphasized the need for more immersive and flexible environments that support active student engagement and higher-order problem solving \cite{Hundhausen2002}.

\subsection{Minecraft Mechanics and Educational Application}
To bypass the technical limitations of custom platforms, educators increasingly utilize stable, commercial off-the-shelf sandbox games. \textit{Minecraft} has emerged as a premier educational tool, with educators frequently taking on the role of "builders" to design custom, interactive laboratories that target STEM skills, creativity, and critical thinking \cite{BarEl2020}. The platform's efficacy extends into higher education, where it is successfully used to teach fundamental programming literacy to university students \cite{AlJanah}. Furthermore, the environment is robust enough to teach advanced computer science paradigms, including object-oriented, event-driven, and parallel programming, by allowing students to interact physically with code-generated structures \cite{Worasait2025}.

The native voxel-based grid of the game naturally maps to mathematical instruction, forcing learners to engage in spatial reasoning and geometric problem-solving \cite{MinecraftEducation}. The platform's adaptability also allows for the design of specialized Human-Computer Interaction (HCI) frameworks that accommodate progressive complexity, ensuring that the interface and learning curve can be tailored to diverse infrastructural and learner needs \cite{Mohd2025}. Finally, allowing students to "remix" or dynamically modify the rules and terrain of their digital environment casts them in the "protagonist role," shifting them from passive consumers of educational content to active, empowered co-creators of their learning experience \cite{Weixelbraun2024}.

\subsection{Proposed Synthesis}
Our proposed system synthesizes these pedagogical necessities. By utilizing the stable, immersive 3D environment of \textit{Minecraft}, we provide the complex graph visualizations that students demand. By tasking students with physically altering terrain to manipulate pathfinding algorithms, the system casts them as active protagonists solving spatial puzzles, effectively addressing the cognitive load and engagement challenges of theoretical computer science education.

\begin{figure}[!h]
  \centering
  \includegraphics[width=7.5cm]{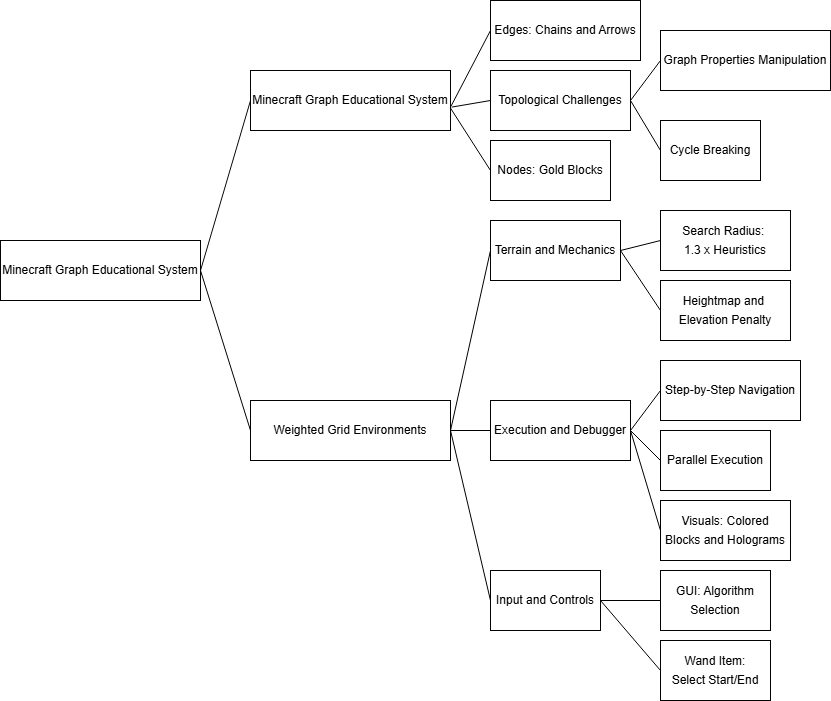}
  \caption{Technical overview of the system design}
  \label{fig:system_design}
\end{figure}

\section{\uppercase{System Design}}
\label{sec:system}
\noindent The proposed system (fig. \ref{fig:system_design})is implemented as a modification for \textit{Minecraft}, built upon the Fabric API. It consists of three distinct interaction modes designed to scaffold the learner's understanding through varying degrees of abstraction.

The system moves from the concrete to the abstract: first, by utilizing the game's native voxel grid to ground pathfinding in familiar spatial reasoning; second, by introducing "Sky Graphs" that isolate topological relationships in a 3D environment; and finally, by providing theoretical reinforcement through embedded pedagogical materials. The purpose of these layers is to make the student transition from a passive observer to an active participant who can manipulate the environment while also understanding the underlying mathematical logic in real-time.

\subsection{Visualizing Traversal}
\begin{figure}[!h]
\centering
\includegraphics[width=7.5cm]{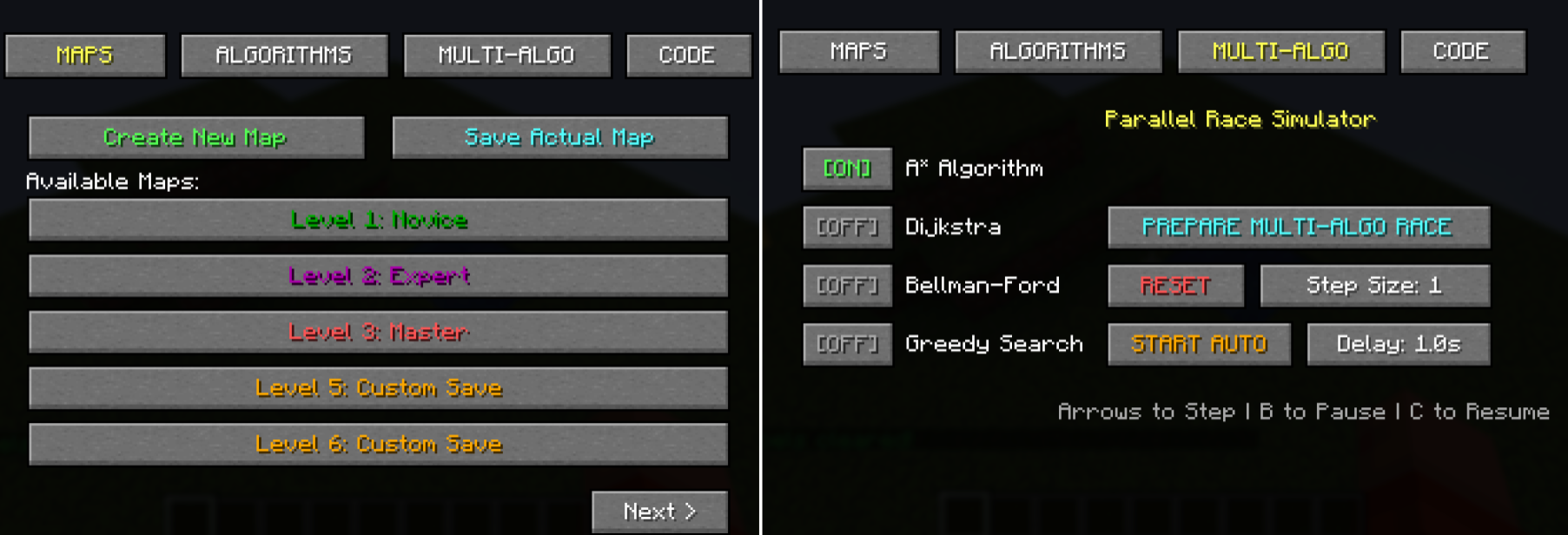}
\caption{GUI for map selection/creation/saving (left) and algorithm selection/debugging and debugger settings (right).}
\label{fig:gui}
\end{figure}

To facilitate ease of use and keep students focused on the algorithmic concepts rather than interface navigation, we implemented a custom in-game item (the "wand") that allows users to select the start and end nodes simply by clicking on the respective blocks. Algorithm selection is currently handled via a dedicated GUI (Figure \ref{fig:gui}).

During execution, the system visually distinguishes between the "visited" and the "frontier" queue by applying contrasting colors to the blocks. This visual separation is crucial, as it allows students to observe the distinct expansion of BFS in relation to the targeted, heuristic-driven pathing of A*.
To accommodate different learning paces, the execution speed of the algorithms can be artificially slowed down. Also, a debugger tool is being implemented to give students absolute control over the execution of the algorithm, featuring the classic "next", "back", "continue", and "break" functions. This step-by-step interactivity allows learners to trace the algorithm's decision-making process at their own speed, reinforcing comprehension.

The system also supports the parallel execution of multiple algorithms. When run concurrently, each algorithm's traversal path is assigned a unique color, allowing students to directly compare their spatial behaviors within the exact same environment. Upon completion, comparison metrics such as the total number of visited nodes, execution speed and path cost are printed directly into the in-game chat. The immediate feedback bridges the gap between spatial visualization and theoretical algorithmic efficiency.

\begin{figure}[!h]
\centering
\includegraphics[width=7.5cm]{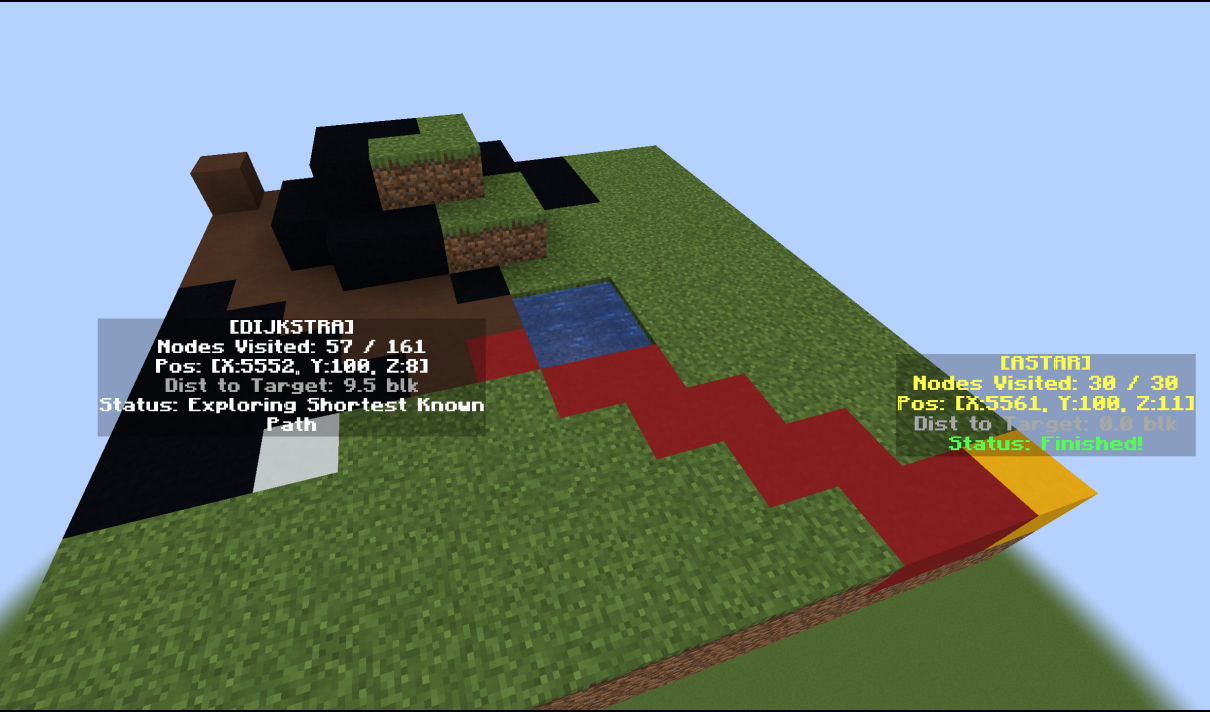}
\caption{Algorithm debugger example. Two algorithms are ran in parallel: Dijkstra (black visited nodes with white current node) and A* (red visited nodes with yellow current node). Some debug info is shown above each of algorithm's current node.}
\label{fig:debugger}
\end{figure}

\subsubsection{The Weighted Grid Environment}
\begin{figure}[!h]
  \centering
  \includegraphics[width=7.5cm]{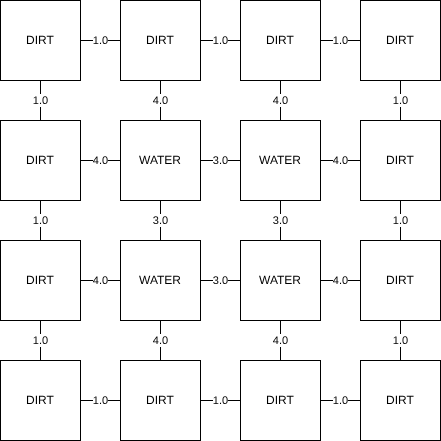}
  \caption{Modelling the Minecraft world as a graph (represented with 4-way edges instead of 8 for clarity)}
  \label{fig:diagram}
\end{figure}

\noindent The core module abstracts a designated area of the \textit{Minecraft} world into a grid of N×M blocks, representing a mathematical graph where each block functions as a discrete node. Edges are established between adjacent blocks, allowing for 8-way horizontal traversal (fig. \ref{fig:diagram}).

The weight of each edge is dynamically calculated based on the material composition of the two connected nodes. For example, the traversal cost between two standard dirt blocks is set to 1.0, whereas transitioning from a dirt block to a water block carries a penalty weight of 4.0, heavily discouraging the pathfinding algorithm from utilizing that route. The system initializes with a predefined lookup table for all possible block-to-block combinations, which educators can easily modify via configuration files to suit specific lesson plans.

The system leverages a heightmap to process the terrain. All relevant blocks are loaded into memory exclusively at the beginning of the execution. To constrain the computational scope, the operational range of the algorithm is restricted to a search circle with a radius equal to $1.30 \times \text{heuristic}(\text{start}, \text{final\_stop})$. The purpose of the constrained search is to balance accuracy with performance. Through iterative trial and error, it was found that smaller radii created "blind spots" where algorithms failed to navigate around large obstacles. Conversely, values exceeding 1.30 increased memory consumption without improving path quality. This multiplier serves as a safety buffer, ensuring the system captures sufficient terrain data while maintaining smooth server performance. 

Because the algorithm evaluates a static snapshot of the loaded world state before executing the pathfinding algorithms, any future terrain modifications, such as placing or destroying blocks, will not be taken into consideration when calculating the route. These changes require the algorithm to be re-run to compute the new path. Furthermore, the system accounts for verticality; elevation changes of a single block between adjacent nodes introduce additional customizable weight penalties to accurately simulate the traversal delay of jumping.

\subsubsection{Pathfinding Challenges}
A key gameplay mode requires students to modify the terrain to force an algorithm to change its path. For example, "Place enough \textit{Soul Sand} to make the straight path more expensive than the longer, curved path".\footnote{\textit{Soul Sand} - block type that slows down any entity that steps on it (granting it a higher cost)} A playground mode will also be provided, which allows the execution of the pathfinding algorithms on any (modifiable) terrain, with start and end points chosen by the player. Other challenges leveraging this visualization mechanic include prompting the player to pick the start and end positions for a certain cost of the shortest road and predicting the next step in an algorithm, given a set of circumstances.

\begin{figure}[!h]
\centering
\includegraphics[width=7.5cm]{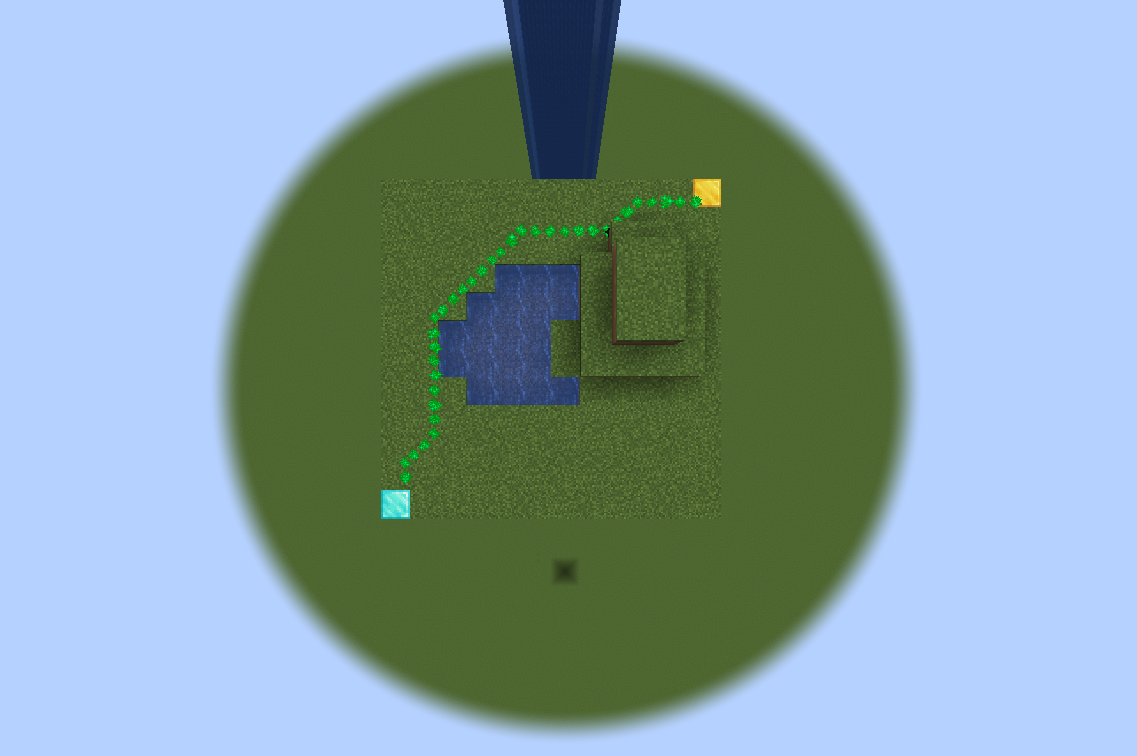}
\caption{Example of pathfinding on a 12x12 map, featuring patches of land on water and height differences in the middle.}
\label{fig:pathfinding}
\end{figure}

\subsubsection{Algorithm Debugger and Execution Control}
\noindent To provide granular control over the learning experience, the system features an integrated algorithm debugger operated via standard keyboard inputs. Students can pause the execution using the 'b' (break) key and resume automatic execution with the 'c' (continue) key. During the continuous playback mode, the execution speed is fully adjustable to accommodate different observation paces. 

For step-by-step analysis, users navigate the algorithm's execution flow using the left and right arrow keys to move backward and forward. The backward stepping functionality is achieved by maintaining a chronological list of all previously visited nodes in memory, allowing the system to revert to earlier states accurately. When a student steps forward, the game world updates immediately: the newly processed block is highlighted as the "current" node, and the "visited" queue expands, with each state represented by distinct block colors. 

Furthermore, pausing the algorithm exposes important underlying metrics to the student. By inspecting the current node, learners can view its path cost, the distance to the final goal, the current status of the algorithm and the number of visited nodes up to that point.

\subsection{Sky-Graphs/Topological Interaction}
A slightly more abstract layer is made up of mini-games which create a temporary "sky-graph" (meaning a floating set of blocks and arrows of chains, depending on whether the graph is directed or not). These mini-games are suited for more traditional problems, such as: adding or removing edges so that a graph acquires a certain property, finding critical nodes/edges, crafting a graph that meets some given requirements from scratch etc. The advantage this visualization method has over pen and paper methods is it can optionally expand to the third dimension, allowing for more interesting layouts and puzzles.

\textbf{Puzzle Example - The Cycle Breaker:} The system generates an undirected graph with a cycle. The student must identify the edge creating the cycle and physically destroy the corresponding Chain block to make the graph acyclic, completing the quest.

\begin{figure}[!h]
\centering
\includegraphics[width=7.5cm]{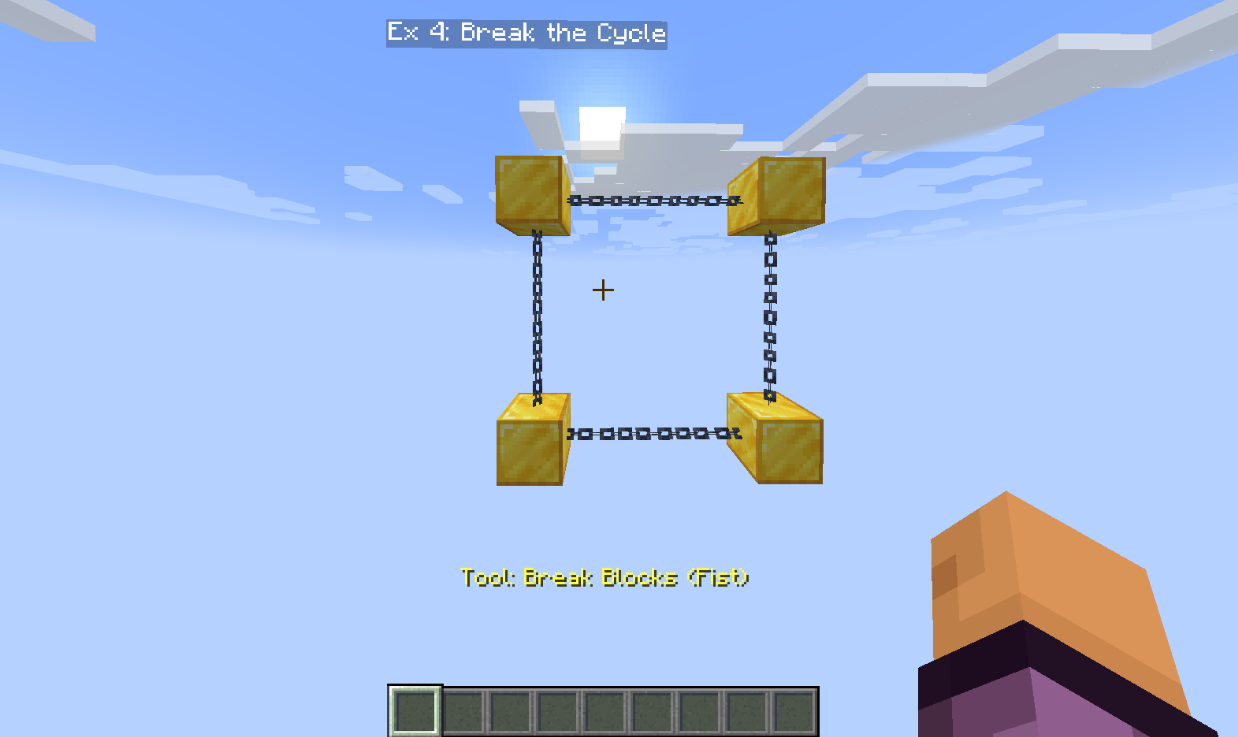}
\caption{Example of Sky Graph: the player must break any one of the chains (edges of the graph) in order to break the cycle and complete the task.}
\label{fig:skygraph}
\end{figure}

\subsection{In-Game Educational Materials}

The most theoretical layer of the system is realized through custom in-game Book items, which serve as a pedagogical bridge between traditional instruction and digital experimentation. While the system emphasizes spatial interaction, the inclusion of structured text ensures a smooth transition for students moving from a traditional classroom setting to the digital environment. 

\begin{figure}[!h]
\centering
\includegraphics[width=7.5cm]{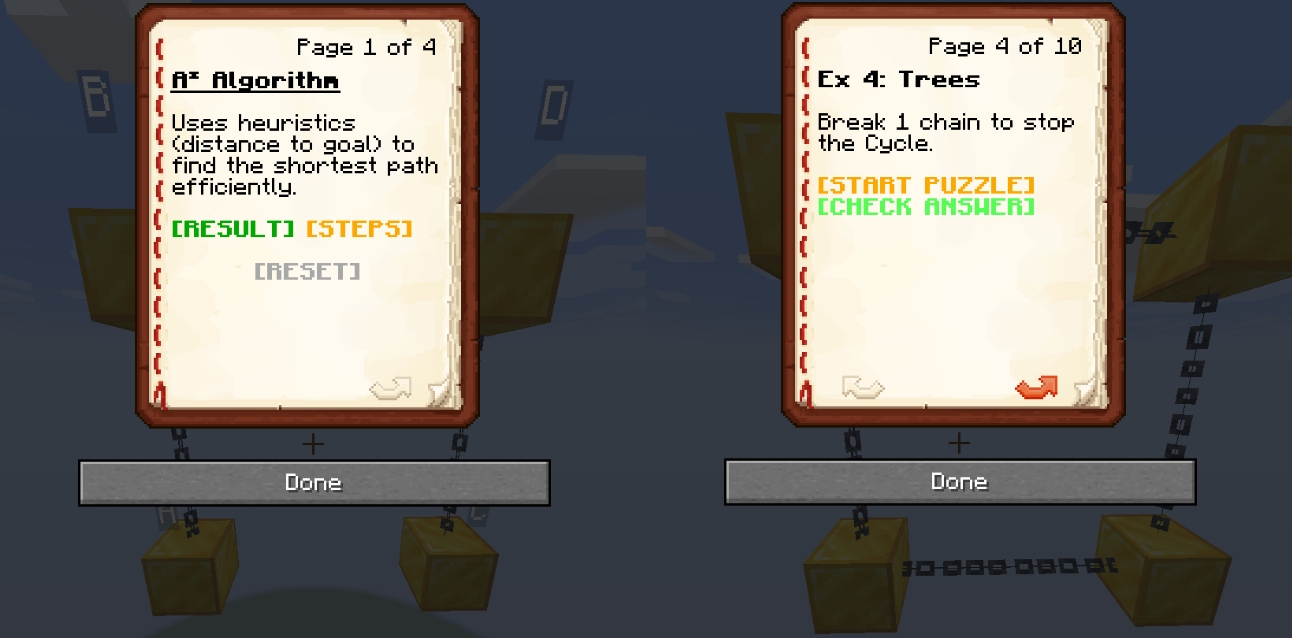}
\caption{Lesson and exercise from a Book}
\label{fig:book}
\end{figure}

Beyond theory delivery, these materials serve as the primary mode of evaluation within the framework. Each book is designed to contain a lesson followed by one or more interactive, multiple-choice quizzes (Figure \ref{fig:book}). These assessments act as the first gate in the learning scaffold: students must demonstrate a baseline understanding of the algorithmic logic before advancing to the high-stakes terrain modification challenges. 

\section{\uppercase{Technical Implementation}}
\label{sec:implementation}

\noindent The modification follows a Client-Server architecture. The "Graph Engine" runs on the server, calculating pathing steps, while the Client handles rendering.

\noindent To ensure server stability, algorithm execution and world modifications are carefully managed across threads. Because \textit{Minecraft}'s \texttt{ServerWorld} is not thread-safe, all visual updates, such as placing concrete blocks or spawning text holograms, are synchronously passed back to the main server tick thread using \texttt{world.getServer().execute()}. Conversely, the auto-play delay loop runs asynchronously via \texttt{CompletableFuture.runAsync}. This allows the use of \texttt{Thread.sleep()} to control execution speed without freezing the server, safely queuing block placement tasks to the main thread tick by tick.

\noindent The architecture leverages the Fabric API for input and event handling. Keyboard inputs for the debugger are integrated into the game's native input system using \texttt{KeyBindingHelper.registerKeyBinding}. The system polls for these key presses at the end of every client frame via \texttt{ClientTickEvents.END\_CLIENT\_TICK.register}, ensuring the render thread remains uninterrupted. For client-server communication, the system bypasses custom networking APIs (such as \texttt{ServerPlayNetworking}). Instead, the client executes hidden chat commands, processed via Brigadier command trees built with \texttt{CommandRegistrationCallback} and \texttt{ClientCommandRegistrationCallback}, to trigger server-side logic.

Visualization states are transmitted to the client as standard block state and entity updates. The system avoids using custom particle packets; instead, the debugger maintains a history of \texttt{BlockChange} objects. During execution or manual stepping, the server physically places colored blocks (e.g., yellow concrete or cyan terracotta) and spawns \texttt{DisplayEntity.TextDisplayEntity} instances for floating holograms. The client processes these exactly as if a player had placed a block or spawned an entity.

Terrain processing utilizes a just-in-time local snapshot approach rather than statically caching heightmap data across server ticks. When an algorithm is prepared for execution, the \texttt{Road\_Manager.scanSurface} method dynamically polls the live world using \texttt{world.getTopY}, generating a fresh, localized map snapshot for that specific run. This inherently captures any recent terrain modifications made by students, effectively detecting changes by reading the live state of the world without requiring complex block-update listeners.

\begin{figure}[!h]
\centering
\includegraphics[width=7.5cm]{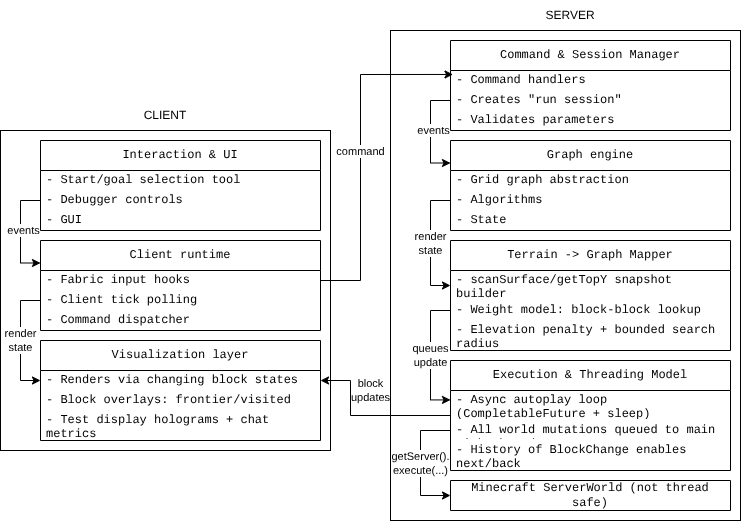}
\caption{Technical architecture overview.}
\label{fig:archoverview}
\end{figure}

\section{\uppercase{Pedagogical Justification}}
\label{sec:pedagogy}
\noindent Although a pilot study is forthcoming, the system design is heavily grounded in Constructionist learning theory and structured scaffolding.

\subsection{Usage Scenario and Hosting Requirements}
\noindent The system is designed for a blended learning environment. Students are initially taught graph theory concepts in a traditional classroom setting without the use of computers. Following this theoretical introduction, students transition to a computer laboratory or an at-home assignment where they launch \textit{Minecraft} to read the in-game lesson books and solve the associated challenges.

Students are instructed to use the "wand" item to experiment and visualize the practical applications of graph traversal algorithms. They are strongly advised to utilize the debugging tool to gain a deeper understanding of the block exploration logic and the decision-making process of the algorithms. Regarding hosting requirements, the modification supports local execution. Students can install and run the system locally in single-player mode.

\subsection{Constructionism and Tangible Artifacts}

\noindent The core tenet of Constructionism is that learners construct mental models most effectively by building tangible, shareable artifacts \cite{Papert2020}. In our system, the artifacts are the modified terrain environments and the constructed 3D Sky Graphs. Students are not just observing algorithms; they are actively shaping the physical environment upon which the algorithms operate.

\subsection{Mapping the Abstract to the Concrete}
\noindent To reduce the cognitive load associated with abstract mathematical structures \cite{Sweller1988}, the system maps discrete graph components to concrete, physical game mechanics:
\begin{itemize}
\item \textbf{Edge Weights:} Increased traversal cost is physically represented by water blocks and elevation changes, mapping a numerical abstraction to a sensory delay.
\item \textbf{Nodes and Edges:} In the topological skygraph mode, nodes are materialized as Gold Blocks. Connections are represented physically using Chains for undirected graphs and Arrows for directed graphs, extending the traditional 2D whiteboard model into a manipulatable 3D space.
\end{itemize}

\subsection{Failure as Learning and Immediate Feedback}
\noindent The system embraces rapid iteration through explicit, immediate feedback mechanisms. In terrain modification exercises, students are tasked with altering the environment to force a pathfinding algorithm to take at least $N$ steps. When the algorithm executes, the path and step count are visualized in real-time. If the threshold $N$ is not met, the visual failure is immediately apparent, encouraging the student to instantly adjust their obstacle placement and retry the execution.

\subsection{Scaffolding and Progressive Complexity}
\noindent To prevent cognitive overload, the system's challenges are structured along a progressively complex learning curve.
\begin{enumerate}
\item \textbf{Familiarization:} The sequence begins with easy quizzes and light theory delivered via in-game \textit{Knowledge Books}. The native, familiar aesthetic of \textit{Minecraft} books draws students in and ensures they acquire the necessary theoretical foundation before attempting physical tasks.
\item \textbf{Topological Manipulation:} Students move to intermediate exercises involving the modification of 3D skygraphs, isolating graph properties from spatial terrain generation.\item \textbf{Applied Pathfinding:} The final tier involves complex pathfinding puzzles set on custom-generated terrain islands, requiring the synthesis of graph theory and spatial reasoning.
\end{enumerate}

\section{\uppercase{Future Work}}
\label{sec:futurework}
\subsection{Advanced Reward System}
\noindent To further incentivize student engagement, future iterations will introduce an advanced in-game reward economy. Successfully completing a challenge will award a predetermined score based strictly on the puzzle's designated difficulty and the fulfillment of the prompt's specific conditions.

\noindent Students will be able to utilize this score to trade with custom in-game \textit{Villager NPCs}.\footnote{\textit{Villager NPC} (non-playable character) - neutral Minecraft entity capable of transactional exchange with the player.} These merchants will offer utility items, such as a "wand" that streamlines the selection of pathfinding start and end nodes, as well as recreational items like speed-enhancing boots and combat weapons.

\noindent These recreational items interface with newly planned casual mini-games, including parkour courses and combat arenas. These mini-games are intentionally disconnected from graph theory, serving as extrinsic rewards and necessary cognitive breaks between intensive learning episodes to prevent mental fatigue and improve information retention.

\subsection{\uppercase{Empirical Classroom Evaluation}}
\label{sec:evaluation}
\noindent To validate the pedagogical efficacy of the proposed system, we aim to conduct a formal pilot study. The evaluation is designed to measure student engagement, cognitive load reduction, and long-term knowledge retention.

\subsubsection{Participants}
\noindent The target demographic consists of students aged 15 to 21, capturing the transition period between secondary and early tertiary education. The primary inclusion criterion is that participants must have no prior formal introduction to graph theory or shortest path algorithms, ensuring an unbiased baseline.

\subsubsection{Study Design and Procedure}
\noindent The evaluation will employ a pre-test and delayed post-test methodology.  Participants will complete an initial assessment to establish baseline knowledge. Following the intervention with the learning system, a delayed post-test will be administered to specifically evaluate long-term knowledge retention rather than immediate, short-term recall.

\subsubsection{Data Collection and Metrics}
\noindent The study will gather quantitative performance data and qualitative feedback to assess the system's impact:
\begin{itemize}
    \item \textbf{Cognitive Load:} The NASA Task Load Index (NASA-TLX) \cite{NASATLX} will be administered post-intervention to quantify perceived cognitive demand and determine if the physicalization of abstract concepts reduces mental strain. 
    \item \textbf{Quantitative Telemetry:} Behavioral data will be extracted directly from the game engine. The primary metric is the number of failed attempts per challenge, which provides insight into the students' iterative problem-solving processes and resilience.
    \item \textbf{Subjective Experience:} Open-ended surveys will be distributed to capture qualitative feedback regarding student engagement, interface usability, and overall satisfaction with the gamified environment.
\end{itemize}

\subsection{Lua Scripting API for Educators}
\noindent To lower the technical barrier for educators, the current Java-based API will be expanded to support Lua integration. Teachers will be able to script custom pathfinding algorithms and traversal logic using lightweight Lua scripts directly within the game environment, completely eliminating the need to modify or recompile the modification's underlying Java codebase.

\section{\uppercase{Conclusions}}
\label{sec:conclusion}
\noindent We have presented the design of a Minecraft-based platform for teaching graph algorithms. By leveraging the game's voxel grid for pathfinding and 3D space for topological graphs, we aim to lower the barrier to entry for complex CS topics. The system physicalizes abstract mathematical structures, mapping numerical edge weights to terrain traversal costs and visualizing execution states through an integrated step-by-step debugger. This constructionist approach transitions learners from passive observers to active protagonists, allowing them to dynamically modify their environment to solve spatial puzzles. Ultimately, the forthcoming empirical classroom evaluation will determine this framework's capacity to reduce cognitive load and improve long-term retention in computer science education.

\section*{Acknowledgments}
The authors used large language model assistants—including Claude (Anthropic), ChatGPT (OpenAI), and Gemini (Google)—to assist with manuscript revision, including improving text clarity, identifying structural issues, and refining the abstract. All content was verified and approved by the authors, who take full responsibility for the accuracy and integrity of this work.

\bibliographystyle{apalike}
{\small
  \bibliography{references}

@misc{DemandSage,
  author       = {{Kumar, Naveen}},
  title        = {How Many People Play Minecraft},
  year         = {2026},
  howpublished = {\url{https://www.demandsage.com/minecraft-statistics}},
  note         = {Accessed: 2026-02-23}
}

@article{Arouri2024,
title={Reluctance of Students to Utilize Virtual Educational Environments in Public Schools: Real World Experiences},
volume={18},
author={Arouri, Yousef M. and Hamaidi, Diala A. and Aldrou’, Islam T. and Noufal, Rana K.},
year={2024}, month={Jan.},
url={https://online-journals.org/index.php/i-jim/article/view/43597},
DOI={10.3991/ijim.v18i01.43597}
}

@inbook{AlJanah,
author = {AlJanah, Salem and Teh, Pin Shen and Tay, Jin and Aiyenitaju, Opeoluwa and Nawaz, Raheel},
year = {2023},
month = {03},
pages = {465-476},
title = {Minecraft as a Tool to Enhance Engagement in Higher Education},
isbn = {978-3-031-19559-4},
doi = {10.1007/978-3-031-19560-0_38}
}

@article{Simonak2014,
author = {Simonak, Slavomir},
year = {2014},
month = {10},
title = {Using algorithm visualizations in computer science education},
volume = {4},
journal = {Open Computer Science},
doi = {10.2478/s13537-014-0215-4}
}

@inbook{Meissner2026,
author = {Mei\ss{}ner, Niklas and Speth, Sandro and Krieger, Niklas and Becker, Steffen},
title = {Boosting Student Motivation through Game-based Learning in Programming Education with Gamify-IT},
year = {2026},
isbn = {9798400722561},
publisher = {Association for Computing Machinery},
address = {New York, NY, USA},
url = {https://doi.org/10.1145/3770762.3772508},
booktitle = {Proceedings of the 57th ACM Technical Symposium on Computer Science Education V.1},
pages = {729–735},
numpages = {7}
}

@inproceedings{Jordaan2025,
author = {Jordaan, Steven and Timm, Nils},
title = {AutomaTutor 2.0: Competitive Game-based Learning of Automata Theory},
year = {2025},
isbn = {9798400719295},
publisher = {Association for Computing Machinery},
address = {New York, NY, USA},
url = {https://doi.org/10.1145/3736181.3747140},
doi = {10.1145/3736181.3747140},
booktitle = {Proceedings of the ACM Global Computing Education Conference 2025 - Volume 1},
pages = {197–203},
numpages = {7},
location = {Gaborone, Botswana},
series = {CompEd 2025}
}

@article{Weixelbraun2024,
author = {Weixelbraun, Petra F. and G\"{o}bl, Barbara and Steinb\"{o}ck, Matthias and Duvivi\'{e}, Mirjam and Kayali, Fares},
title = {Discussing the Protagonist Role of Students in Game-Based Learning},
year = {2024},
issue_date = {October 2024},
publisher = {Association for Computing Machinery},
address = {New York, NY, USA},
volume = {8},
number = {CHI PLAY},
url = {https://doi.org/10.1145/3677065},
doi = {10.1145/3677065},
journal = {Proc. ACM Hum.-Comput. Interact.},
month = oct,
articleno = {300},
numpages = {24}
}

@inproceedings{BarEl2020,
author = {Bar-El, David and E. Ringland, Kathryn},
title = {Crafting Game-Based Learning: An Analysis of Lessons for Minecraft Education Edition},
year = {2020},
isbn = {9781450388078},
publisher = {Association for Computing Machinery},
address = {New York, NY, USA},
url = {https://doi.org/10.1145/3402942.3409788},
doi = {10.1145/3402942.3409788},
booktitle = {Proceedings of the 15th International Conference on the Foundations of Digital Games},
articleno = {90},
numpages = {4},
location = {Bugibba, Malta},
series = {FDG '20}
}

@misc{Mohd2025,
      title={Khelte Khelte Shikhi: A Proposed HCI Framework for Gamified Interactive Learning with Minecraft in Bangladeshi Education Systems}, 
      author={Mohd Ruhul Ameen and Akif Islam and Momen Khandokar Ope},
      year={2025},
      eprint={2510.18385},
      archivePrefix={arXiv},
      primaryClass={cs.HC},
      url={https://arxiv.org/abs/2510.18385}, 
}

@inproceedings{Worasait2025,
  author    = {Suwannik, Worasait},
  title     = {Minecraft: An Engaging Platform to Learn Programming},
  booktitle = {2025 IEEE/ACIS 23rd International Conference on Software 
               Engineering Research, Management and Applications (SERA)},
  year      = {2025},
  publisher = {IEEE}
}

@misc{MinecraftEducation,
  author       = {{Minecraft Education}},
  title        = {Using Minecraft for Mathematics Instruction},
  year         = {2024},
  howpublished = {\url{https://education.minecraft.net/}},
  note         = {Accessed: 2026-02-23}
}

@book{Papert2020,
  title={Mindstorms: Children, Computers, And Powerful Ideas},
  author={Papert, S.A.},
  isbn={9781541675100},
  url={https://books.google.ro/books?id=nDjRDwAAQBAJ},
  year={2020},
  publisher={Basic Books}
}

@article{Sweller1988,
title = {Cognitive load during problem solving: Effects on learning},
journal = {Cognitive Science},
volume = {12},
number = {2},
pages = {257-285},
year = {1988},
issn = {0364-0213},
doi = {https://doi.org/10.1016/0364-0213(88)90023-7},
url = {https://www.sciencedirect.com/science/article/pii/0364021388900237},
author = {John Sweller}
}

@incollection{NASATLX,
title = {Development of NASA-TLX (Task Load Index): Results of Empirical and Theoretical Research},
editor = {Peter A. Hancock and Najmedin Meshkati},
series = {Advances in Psychology},
publisher = {North-Holland},
volume = {52},
pages = {139-183},
year = {1988},
booktitle = {Human Mental Workload},
issn = {0166-4115},
doi = {https://doi.org/10.1016/S0166-4115(08)62386-9},
url = {https://www.sciencedirect.com/science/article/pii/S0166411508623869},
author = {Sandra G. Hart and Lowell E. Staveland}
}

@article{Hundhausen2002,
title = {A Meta-Study of Algorithm Visualization Effectiveness},
journal = {Journal of Visual Languages and Computing},
volume = {13},
number = {3},
pages = {259-290},
year = {2002},
issn = {1045-926X},
doi = {https://doi.org/10.1006/jvlc.2002.0237},
url = {https://www.sciencedirect.com/science/article/pii/S1045926X02902375},
author = {CHRISTOPHER D. HUNDHAUSEN and SARAH A. DOUGLAS and JOHN T. STASKO}
}
}

\end{document}